\documentclass[aip,jap,
12pt,
floatfix,
]{revtex4-1}
\usepackage{amsmath}    

\DeclareMathOperator{\sgn}{sgn}
\DeclareMathOperator{\Sinc}{Sinc}
\usepackage{graphicx}   
\usepackage{bmpsize}
\usepackage[hang]{subfigure}
\newcommand{\jun}{junction }
\newcommand{\juns}{junctions }
\newcommand{\Jos}{Josephson }
\newcommand{\elli}{elliptic }

\begin{document}
\title[R.Monaco \textit{et al.}]{Superelliptic Josephson Tunnel Junctions}
\author{Roberto Monaco}
\affiliation{CNR-ISASI, Institute of Applied Sciences and Intelligent Systems ''E. Caianello'', Comprensorio Olivetti, 80078 Pozzuoli, Italy}
\email[Corresponding author e-mail:$\,$]{$\,$r.monaco@isasi.cnr.it,roberto.monaco@cnr.it}
\date{\today}

\begin{abstract}
\begin{large}
The most important practical characteristic of a Josephson junction is its critical current. The shape of the junction determines the specific form of the magnetic-field dependence of the its Josephson current. Here we address the magnetic diffraction patterns of specially shaped planar Josephson tunnel junctions. We focus on a wide ensemble of generalized ellipses, called superellipses, which retain the second order symmetry. We analyze the implications of this type of isometry and derive the explicit expressions for the threshold curves of superelliptic Josephson junctions. A detailed study is made of their magnetic patterns with emphasis on the rate of decay of the sidelobes amplitudes for large field amplitudes.

\end{large}
\end{abstract}
\maketitle
\section{Introduction}

Any \Jos device is characterized by a maximum zero-voltage d.c. current, $I_c$, called critical current, above which it switches to a finite voltage. How the critical current modulates with an external magnetic induction field, $H$, provides the first quality test of a Josephson junction. It has long been addressed that the critical-current pattern as a function of magnetic field, $I_c(H)$, also called  the magnetic diffraction pattern (MDP), of planar \Jos tunnel \juns (JTJs) drastically depends on the specific shape of the tunneling area \cite{barone}. For a rectangular junction this results in the well-known \textit{Fraunhofer pattern}, whose modulation envelope falls off as $1/H$ with the magnetic field along an axis; in contrast, the \textit{Airy pattern} derived for a circular JTJ \cite{barone} at large fields decreases as $1/H^{3/2}$. Peterson and Ekin \cite{ekin} showed that the Airy pattern also apply to elliptic \Jos \juns. It was also pointed out that when the field is along the principal axis of a barrier shaped as a biconvex (converging) lens, the pattern generally falls-off (more) steeply and is sensitive to the detail of the shape \cite{peterson91}. The study of the shape dependence of the MDP is not only a rich and fascinating subject in its own right, but a confluence of several relevant application issues. 

The transport critical current density, $J_c$, in bulk high-T$_c$ material decreases by about
two orders of magnitude in an applied magnetic field as low as $10mT$ \cite{Dew-Hughes}. The initial rapid drop in $J_c$ with field is due to many Josephson weak links (insulating or normal barriers characterized by the Josephson $\sin \phi$ relation) between grains being progressively switched off as the field is increased. The geometry of the contact areas between two grains and their orientation are relevant parameters for the texturing of high-$J_c$ materials.


Furthermore, in some applications of JTJs (such as SIS mixing and X-ray detection) the modulation sidelobes of the MDP are required to be as small as possible. So far, several junction shapes were considered and and their critical-current patterns explicitly calculated. Besides the well known \textit{diamond} geometry (square junction with the field along the diagonal), in which the critical-current diffraction pattern falls off with the field as $1/H^2$, various other non-common shapes were considered. Among these, junctions with \textit{quartic} geometry \cite{peterson91} were demonstrated \cite{gijsbertsen95} to be have the most rapid $I_c$ suppression compared with other known geometries, including the (truncated) normal distribution shape \cite{kikuchi00}, with the same transverse dimension. Irregular quadrangle junctions were also considered by Nappi \textit{et al.} \cite{nappi96}.

The milestone works \cite{peterson91,gijsbertsen95,kikuchi00,nappi96} which allowed significant advances in the understanding of the geometrical properties of the MDP just considered a selection of appealing shapes with the magnetic field applied in a preferential direction. However, the number of interesting small-junction shapes so far considered is far from being exhaustive. The aim of this paper is to study JTJs delimited by a class of plane curves, called \textit{superellipses}, retaining the symmetry and the geometric features of major and minor axes of the ellipses, but having a different overall shape. We investigate the magnetic properties of the superelliptic junctions, analyse the modification of their critical-current patterns and show that, for a proper choice of the shape, the asymptotic behavior at infinity of the MDP follows a power law $1/H^n$ with an arbitrarily large scaling exponent.


%


The article is organized as follows: in Sec.II we provide the mathematical representation of superellipses and then extend the definition to \textit{generalized} and \textit{hybrid} superellipses. In Sec.III the basics of the theory of the diffraction patterns of small Josephson junctions in magnetic field are briefly reviewed; the formalism of the junction characteristic function for point symmetric geometries is also introduced. In Sec.IV an analytical expression of the characteristic function of a generic superelliptic junction is derived. Sec.V we provide and discuss the closed form expressions of the characteristic function for superelliptic junctions having several configurations worth noting. In Sec. VI a similar analysis is carried for hybrid superelliptic junctions with a particular emphasis on the issue of sidelobe suppression. Finally, the main points addressed in the article are summarized in Sec.VII. 

\section{The superellipses}

\begin{figure}[t]
\centering
\subfigure[ ]{\includegraphics[width=7cm]{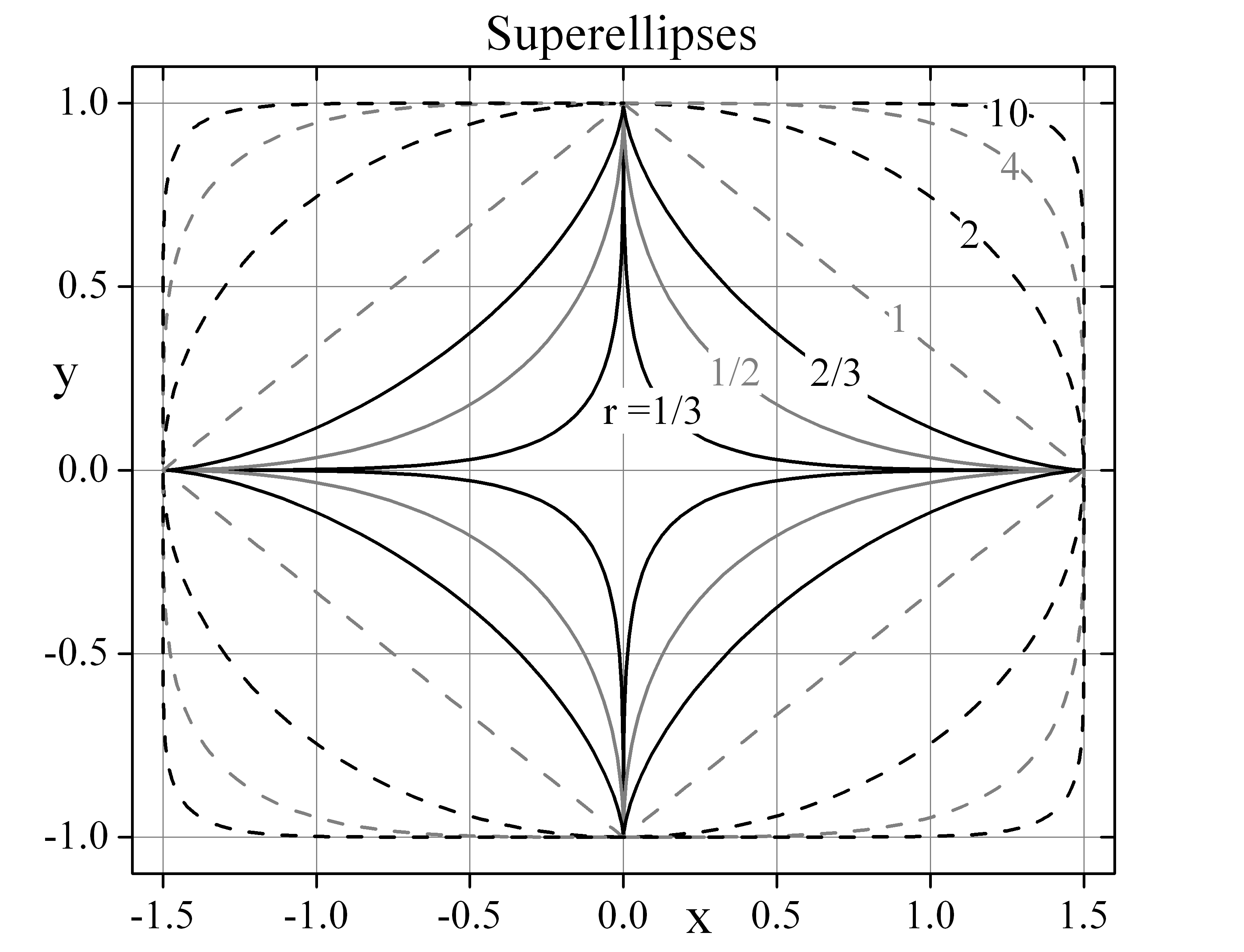}}
\subfigure[ ]{\includegraphics[width=7cm]{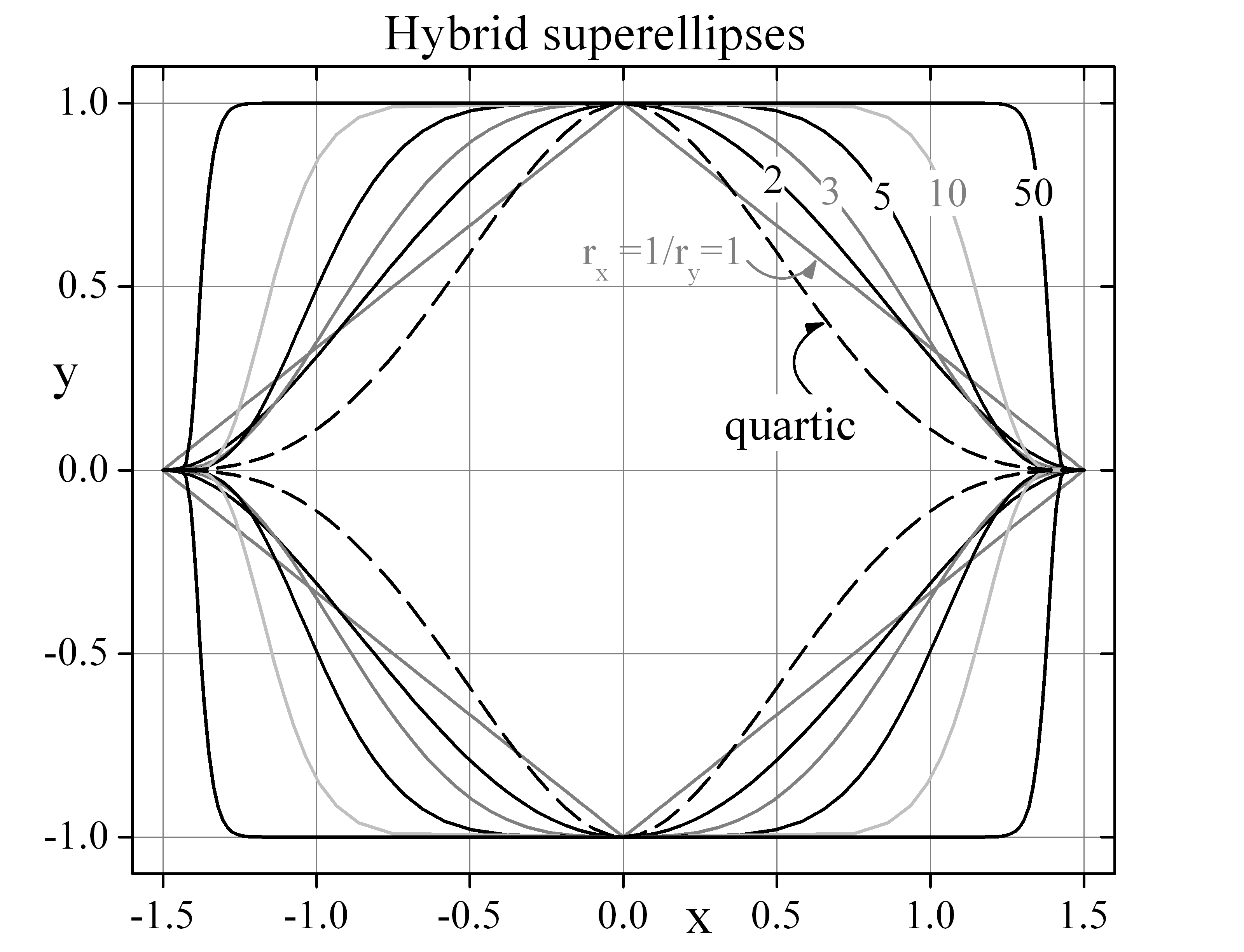}}
\caption{a) Superelliptic curves as in Eq.(\ref{pedal}) for $a=1.5$, $b=1$ and $r=1/3$, $1/2$, $2/3$, $1$, $2$, $4$, and $10$ going from inside to outside. b) Hybrid superelliptic curves as in Eq.(\ref{pedal2}) for $a=1.5$, $b=1$ and  $r_x=1/r_y=1$, $2$, $3$, $5$, $10$ and $50$ going from inside to outside. The dashed line shows the shape of the quartic junction in Eq.(\ref{quartic}).}
\label{draw}
\end{figure}

\noindent The superellipses were first studied by the French mathematician Gabriel Lam\`e in the early 19th century in the midst of his exploration of curvilinear coordinates. In two dimensions the superellipse (or Lam\`e curve) centered at the origin is defined by the implicit equation:
\vskip -8pt
\begin{equation}
\label{pedal}
\left|\frac{x}{a} \right|^r+\left| \frac{y}{b} \right|^r=1
\end{equation}

\noindent with the semi-axes $a$ anb $b$ and the exponent $r$ positive numbers. Solving Eq.(\ref{pedal}) for $y$ gives $y=\pm b \sqrt[r]{1- \left|x/a\right|^r }$. Indicating with $\sgn$ the signum function, the superellipse is described by the parametric equations:
\vskip -8pt
$$
\begin{cases}
x(\tau)=a \left|\sin\tau \right|^{\frac{2}{r}} \sgn\, \sin\tau \\
y(\tau)=b \left|\cos\tau \right|^{\frac{2}{r}} \sgn\, \cos\tau, \end{cases} 
$$

\noindent where $\tau$ is a parameter measured clockwise from the positive $Y$-axis, not to be confused with the polar angle $\theta$ defined as $\theta \equiv \text{ArcTan}\, x/y=\text{ArcTan}\,\tan^{2/r} \tau$. The corresponding polar equation with coordinates $(\rho,\tau)$ is $\rho=\left(\left|\frac{\sin\tau}{a} \right|^r+\left|\frac{\cos\tau}{b} \right|^r\right)^{-1/r}$.
\noindent The Lam\`e curves intersect the $X$-axis at $\pm a$ and the $Y$-axis at $\pm b$, i.e., the constants $2a$ and $2b$ define the figure width and height. If $a>b$ ($a<b$) we talk of an horizontally (vertically) long or oblate (prolate) superellipse; a superellipse may be called a \textit{supercircle} when the aspect ratio $a/b$ equals $1$. The value of the exponent $r$ determines the shape of the curve. As $r$ approaches zero the curve degenerates to two straight crossed segments along the axes; increasing the exponent from $0$ to $\infty$, the curves fill up the area enclosed in a axis-parallel rectangle of width $2a$ and height $2b$. The superellipses with principal semi-axes $a=1.5$ and $b=1$ are drawn in Figure~\ref{draw}(a) for $r = 1/3$, $1/2$, $2/3$, $1$, $2$, $4$ and $10$, going from inside to outside. More specifically, a superellipse is called rectellipse when $r=4$ (or squircle if $a=b$), natural superellipse when $r=e$, ordinary ellipse when $r=2$ (or circle if $a=b$), rhombus when $r=1$ (or diamond if $a=b$), and astroid when $r=2/3$. We observe that for $r<1$ a superellipse is a pinched rhombus or a four-armed star with concave (inwards-curved) sides; for $r = 1/2$, in particular, each of the four arcs is a segment of a parabola. For $1<r<2$ the curve looks like a rhombus with those same corners but with convex (outwards-curved) sides, also called hypo-ellipse. For $r>2$ the curve looks like a rectangle with rounded corners (hyper-ellipse) and becomes a rectangle in the limit $r\to \infty$.

\noindent The superellipses have two perpendicular axes of symmetry whose intercept point is called symmetry center. Therefore, they have a rotational symmetry of order $2$, i.e., they are invariant under a rotation of $360/2=180$ degrees around the symmetry point; such plane figures are also called \textit{point symmetric}. For supercircles the symmetry order becomes $4$.

\subsection{The generalized and hybrid superellipses}

\noindent A generalization of the superellipses is achieved by letting the two exponents in Eq.(\ref{pedal}) to be different, namely,	the \textit{generalized} superellipse centered at the origin is defined by the Cartesian equation:
\vskip -8pt
\begin{equation}
\label{pedal2}
\left|\frac{x}{a} \right|^{r_x}+\left| \frac{y}{b} \right|^{r_y}=1.
\end{equation}

\noindent with $a$, $b$, $r_x$ and $r_y$ positive numbers. Eq.(\ref{pedal2}) may be described parametrically by:
\vskip -8pt
$$
\begin{cases}
x(\tau)=a \left|\sin\tau \right|^{\frac{2}{r_x}} \sgn\, \sin\tau \\
y(\tau)=b \left|\cos\tau \right|^{\frac{2}{r_y}} \sgn\, \cos\tau; \end{cases} 
$$

\noindent the polar representation does not exist for $r_x\neq r_y$. For our purposes, particularly interesting is the case of \textit{hybrid} superellipses defined as (generalized) superellipses having reciprocal exponents $r_x=1/r_y$. The hybrid superellipses with principal semi-axes $a=1.5$ and $b=1$ are drawn by solid lines in Figure~\ref{draw}(b) for $r_x=1/r_y = 1$, $2$, $3$, $5$, $10$ and $50$, going from inside to outside. As the exponent increases the superellipses tend to a rectangular shape with a less and less appreciable pinching on the vertical sides. It is worth noting that even though some superellipses can look as though they have straight horizontal sides joined by curves, they are actually curved all the way around with a calculable and finite center of curvature. For the sake of comparison, the dashed line in Figure~\ref{draw}(b) shows the \textit{quartic} shape with zero slope at the ends proposed in Ref. \cite{peterson91} and defined by the quartic polynomial:

\vskip -8pt
\begin{equation}
y(x)=\pm b\left(1-\left|\frac{x}{a}\right|\right)^2 \left[1 + 2 \left|\frac{x}{a}\right| - 3 \left(\frac{x}{a}\right)^2\right].
\label{quartic}
\end{equation}

\section{Small JTJs in an external magnetic field}

In Josephson's original description the quantum mechanical phase difference, $\phi$, across the barrier of a generic two-dimensional planar \Jos tunnel \jun is related to the magnetic field, ${\bf H}$, inside the barrier through \cite{brian}:
\vskip -8pt
\begin{equation}
\label{gra}{\bf \nabla} \phi =
\kappa{\bf H}\times {\bf u}_z ,
\end{equation}
\noindent in which ${\bf u}_z$ is a unit vector orthogonal to the \jun plane and $\kappa\equiv 2\pi\mu_0 d_m/\Phi_0$, where $\Phi_0$ is the magnetic flux quantum, $\mu_0$ the vacuum permeability, and $d_m$ the \jun \textit{magnetic} penetration depth \cite{wei,SUST13a}. The external field ${\bf H}$, in general, is given by the sum of an externally applied field and the self-field generated by the current flowing in the junction. If the junction dimensions are smaller than the \Jos penetration length, the self-magnetic field is negligible, as has been first shown by Owen and Scalapino \cite{owen} for a rectangular JTJ . Henceforth, for \textit{electrically small} JTJs considered in this paper the phase spatial dependence is obtained by integrating Eq.(\ref{gra}), regardless of the geometry of the current-carrying electrodes \cite{PRB12}; in Cartesian coordinates, for an in-plane magnetic field applied at an arbitrary direction, ${\bf {H}}\equiv (H_x,H_y)$, it is: 
\vskip -8pt
\begin{equation}
\phi(x,y,H_x,H_y,\phi_0)=\kappa (H_y x - H_x y)+\phi_0,
\label{small}
\end{equation}


\noindent where $\phi_0$ is an integration constant. Eq.(\ref{small}) implicitly assume that the junction electrodes are free from permanent circulating current, that is either the electrodes are simply-connected or one or both electrodes are doubly (or multiply) connected with no magnetic flux trapped in the hole \cite{SUST13a}. The tunneling current flows in the $Z$-direction and the local density of the \Jos current is \cite{brian}:
\vskip -8pt
\begin{equation}
\label{jj}J_J(x,y,H_x,H_y,\phi_0) =J_c \sin \phi(x,y,H_x,H_y,\phi_0),
\end{equation}

\noindent where $J_c$ is the maximum \Jos current density. The \Jos current, $I_J$, through the barrier is obtained integrating Eq.(\ref{jj}) over the junction surface, $S$:
\vskip -8pt
\begin{equation}
\label{IJ}
I_J(H_x,H_y,\phi_0)=\int_S J_J\,dS = J_c \int_S \sin [\kappa (H_y x - H_x y)+\phi_0]\,dS.
\end{equation}

\noindent where $J_c$ is assumed to be uniform over the junction area. The \jun critical current, $I_c$, is defined as the largest possible \Jos current, namely,:
\vskip -8pt
\begin{equation}
\label{Ic}
I_c(H_x,H_y)= \max_{\phi_0} I_J(H_x,H_y,\phi_0),
\end{equation}

\noindent It follows that the zero field critical current, $I_0\equiv I_c(H=0)$, is given by the product of the maximum \Jos current density and the tunneling area, $A_S=\int_S dS$, namely, $I_0= J_c A_S$.

%

\subsection{Point symmetric JTJs}

\noindent Also the generalized superellipses are invariant upon reflections in the two perpendicular lines of symmetry, that is to say, they are point symmetric plane figures. It has been recently reported \cite{PhysC16b} that for a point symmetric JTJ, in force of the sine function oddity, the surface integral over its surface $S$ of the term $\sin[\kappa H_y x- H_x y)]$ is automatically zero; therefore, the $\phi_0$-dependence of the \Jos current in Eq.(\ref{IJ}) is simply sinusoidal, for any field direction:
\vskip -8pt
\begin{equation}
\label{IJ1}
I_J(H_x,H_y,\phi_0)=I_0\, \mathcal{F}_S(H_x,H_y) \sin \phi_0,
\end{equation}

\noindent where we defined the characteristic shape-dependent function:
\vskip -8pt
\begin{equation}
\label{mathcalF}
\mathcal{F}_S(H_x,H_y)\equiv \frac{1}{A_S}\int_S \cos\left[\kappa (H_y x-H_x y)\right]\, dS.
\end{equation}

\noindent At last, in force of Eq.(\ref{Ic}), the MDP of a point symmetric junction can be simply expressed as:
\vskip -8pt
\begin{equation}
\label{IcS}
I_c(H_x,H_y)= I_0 \left|\mathcal{F}_S(H_x, H_y)\right|.
\end{equation}

\noindent By definition, $\mathcal{F}_S$ is an area-independent function such that $|\mathcal{F}_S (H=0)|=1$. In different words, the absolute value of the characteristic function yields the the normalized critical current, $i_c(H_x,H_y)\equiv I_c(H_x,H_y)/I_0= \left|\mathcal{F}_S(H_x,H_y)\right|$.

\section{The Magnetic Diffraction Patterns}

In this section we will derive 
an analytical expression for the characteristic function of a planar JTJ whose barrier is delimited by a (generalized) superellipse when the magnetic field is applied along one of the symmetry lines. In force of the symmetry properties of the superellipses, the integral in Eq.(\ref{mathcalF}) can be computed over just the I and II quadrants, where $x\geq0$:
\vskip -8pt
\begin{equation}
\label{Fr}
\mathcal{F}_{S}(H_x,H_y)=\frac{2}{A_S}\int_{I+II} \cos\left[\kappa (H_y x- H_x y)\right]\, dx dy=
\end{equation}
\vskip -8pt
$$=\frac{2}{A_S}\int_{0}^a dx \int_{-y(x)}^{y(x)} dy\, \cos\left[\kappa (H_y x- H_x y)\right].$$

\noindent From Eq.(\ref{pedal2}) with $x\geq0$, we get $y(x)=\pm b [1- (x/a)^{r_x} ]^{1/r_y}$. In the following equations, $\mathcal{F}_{r_x,r_y}$ is the characteristic area-independent function of all the generalized superellipses of degrees $r_x$ and $r_y$. In addition, we will make use of the normalized magnetic field components $h_x\equiv \kappa H_x b$ and $h_y\equiv \kappa H_y a$. Furthermore, the area $A_{r_x,r_y}$ of a generalized superellipse of degrees $r_x$ and $r_y$ is given by \cite{wolfram}:
\vskip -8pt
\begin{equation}
A_{r_x,r_y}=4ab\frac{\Gamma\left(1+\frac{1}{r_x}\right)\Gamma\left(1+\frac{1}{r_y}\right)}{\Gamma\left(1+\frac{1}{r_x}+\frac{1}{r_y}\right)};
\label{areaGen}
\nonumber
\end{equation}
\vskip -4pt
\noindent setting $r_x=r_y=r$, for an ordinary superellipse of degree $r$ the last expression reduces to:
\vskip -8pt
\begin{equation}
A_r=4ab\frac{\Gamma^2\left(1+\frac{1}{r}\right)}{\Gamma\left(1+\frac{2}{r}\right)}=4ab  \frac{\sqrt{\pi} \Gamma\left(1+\frac{1}{r}\right)}{\sqrt[r]{4} \Gamma\left(\frac{1}{2}+\frac{1}{r}\right)}.
\label{area}
\nonumber
\end{equation}
\vskip -4pt

\noindent We first consider the particular case of a magnetic field applied along the $Y$-axis. Introducing the normalized variables $x'\equiv x/a=\sin^{2/r_x}\tau$ and $y'\equiv y/a=\cos^{2/r_y}\tau$, Eq.(\ref{Fr}) with $h_x=0$ can be rewritten as:
\vskip -8pt
$$\mathcal{F}_{r_x,r_y}(h_x=0,h_y)=\frac{4ab}{A_{r_x,r_y}}\int_{0}^1 y'(x')\cos\left(h_y\, x'\right)dx'=$$

\vskip -8pt
\begin{equation}
\label{Fr0Gen}
=\frac{8ab}{r_x A_{r_x,r_y}}\int_0^{\pi/2} \sin^{\frac{2}{r_x}-1}\tau \cos^{\frac{2}{r_t}+1}\tau \cos\left(h_y \sin^{\frac{2}{r_x}}\tau \right)d\tau,
\end{equation}

\noindent being $dx'=\frac{2}{r_x}\sin^{\frac{2}{r_x}-1}\tau \cos\tau d\tau$. For any other field direction, i.e., assuming $h_x\neq 0$, in force of the angle-sum trigonometric identities, Eq.(\ref{Fr}) gives:
\vskip -8pt
$$\mathcal{F}_{r_x,r_y}(h_x \neq 0,h_y)=\frac{8ab}{A_{r_x r_y} h_x}\int_{0}^1 \!\!\! \cos(h_y\, x')\sin(h_x\, y')\, dx'=
$$
\vskip -8pt
\begin{equation}
\label{Frnot0Gen}=\frac{16ab}{r_x A_{r_x,r_y} h_x}\int_{0}^{\pi/2} \!\!\! \cos(h_y \sin^{\frac{2}{r_x}}\tau)\sin(h_x \cos^{\frac{2}{r_y}}\tau)\sin^{\frac{2}{r_x}-1}\tau \cos\tau \, d\tau.
\end{equation}

\noindent For the specific case of $h_y=0$, the above expression reduces to:
\vskip -8pt
\begin{equation}
\label{FrPimezziGen}
\mathcal{F}_{r_x,r_y}(h_x\neq0, h_y=0)=\pm \frac{16ab}{r_x A_{r_x,r_y} h_x}\int_{0}^{\pi/2} \!\!\! \sin(h_x \cos^{\frac{2}{r_y}}\tau)\sin^{\frac{2}{r_x}-1}\tau \cos\tau \, d\tau.
\end{equation}

\noindent We observe that the ratio $ab/A_{r_x,r_y}$ appearing in the previous equations is independent of the dimensions of the superellipse, therefore, as expected, the characteristic function profile only depends on the superellipse shape through the exponents $r_x$ and $r_y$ (the superellipse dimensions only affect the scaling of the magnetic field normalization).

\subsection{Generalized hypergeometric functions}

\noindent For rational $r_x$ and $r_y$, the computation of the the definite integrals in Eqs.(\ref{Fr0Gen})-(\ref{FrPimezziGen})is often quite involved and the superellipse's characteristic functions are expressed in terms of generalized hypergeometric functions, provided that the magnetic field is directed along one of the symmetry axes. The generalized hypergeometric function, $\,_pF_q$, of order $p,q$ is defined as follows:
\vskip -8pt
$$\,_pF_q(a;b;z)=\sum_{k=0}^{\infty}\left[\frac{(a_1)_k (a_2)_k....(a_p)_k}{(b_1)_k (b_2)_k....(b_p)_k} \frac{z^k}{k!}\right],$$

\noindent where the list, $a = [a_1,a_2,...,a_p]$, of numerator coefficients and the list, $b = [b_1,b_2,...,b_q]$, of denominator coefficients are vectors of lengths $p$ and $q$, respectively. $(a_i)_k$ and $(b_j)_k$ are Pochhammer symbols; if $k$ is a positive integer, then $(x)_k = x(x + 1)...(x + k - 1)$. Any hypergeometric function, evaluated at 0, has the value 1. Hypergeometric functions also reduce to other special functions for some parameters; for example, $_0F_0(;;z)=e^z$.

\section{The MDP of super\elli \Jos tunnel junctions}

\noindent We now examine the pattern profile of (ordinary) superellipses of given degree $r$ when the magnetic field is applied perpendicular to the $a$ dimension. The computation of the characteristic area-independent function, $\mathcal{F}_r(h_y)$, is achieved by setting $r_x=r_y=r$ in Eq.(\ref{Fr0Gen}). The definite integrals were analytically solved using a commercial software; the results are reported below for several rational values of the exponent $r$. For non-rational values of the exponent one must resort to numerical integration methods. It is worth noting that, for symmetry reasons, the same results can be obtained from Eq.(\ref{FrPimezziGen}) (valid for $h_y=0$) with $r_x=r_y=r$ and replacing $h_x$ with $b h_y/a$. This alternative path has been used to double-check our findings. Inspection of the explicit expressions shows how fast the envelope of the pattern falls off with field, as well as other details of the pattern.

\subsubsection*{$r=1/4$}
\vskip -8pt
$$\mathcal{F}_{1/4}(h_y)=56 \, _1F_2\left(\frac{5}{8};\frac{3}{2},\frac{13}{8};-\frac{h_y^2}{4}\right)+120 \, _1F_2\left(\frac{7}{8};\frac{3}{2},\frac{15}{8};-\frac{h_y^2}{4}\right)-\frac{140 }{h_y^2}\left[3 \sqrt{\frac{\pi h_y}{2}} 	\mathcal{S}\left( \sqrt{\frac{2 h_y}{\pi }}\right)+\sin ^2\frac{h_y}{2}\right],$$
\vskip -4pt
\noindent where $\mathcal{S}$ is the Fresnel sine integral.

\begin{figure}[tb]
\centering
\subfigure[ ]{\includegraphics[width=7cm]{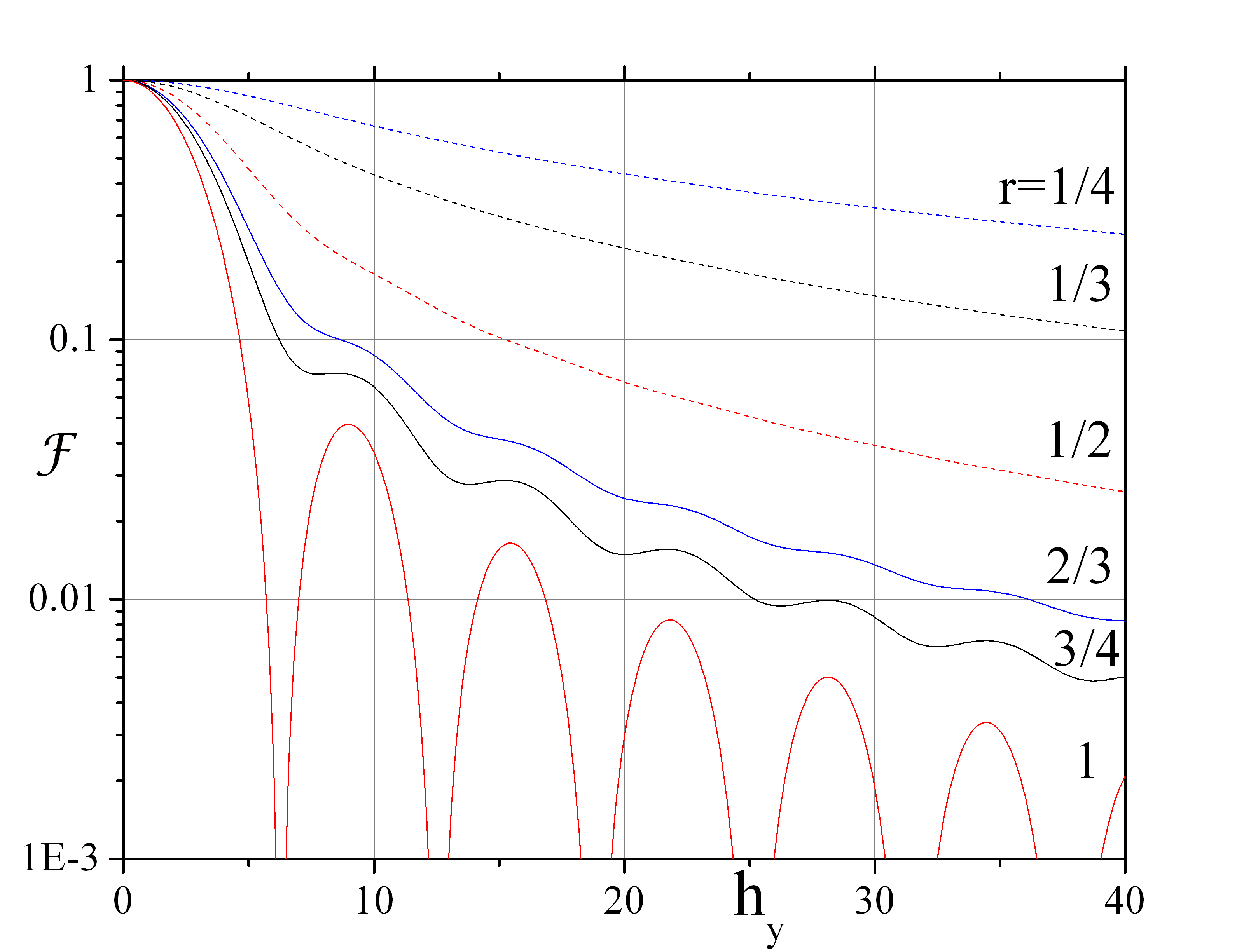}}
\subfigure[ ]{\includegraphics[width=7cm]{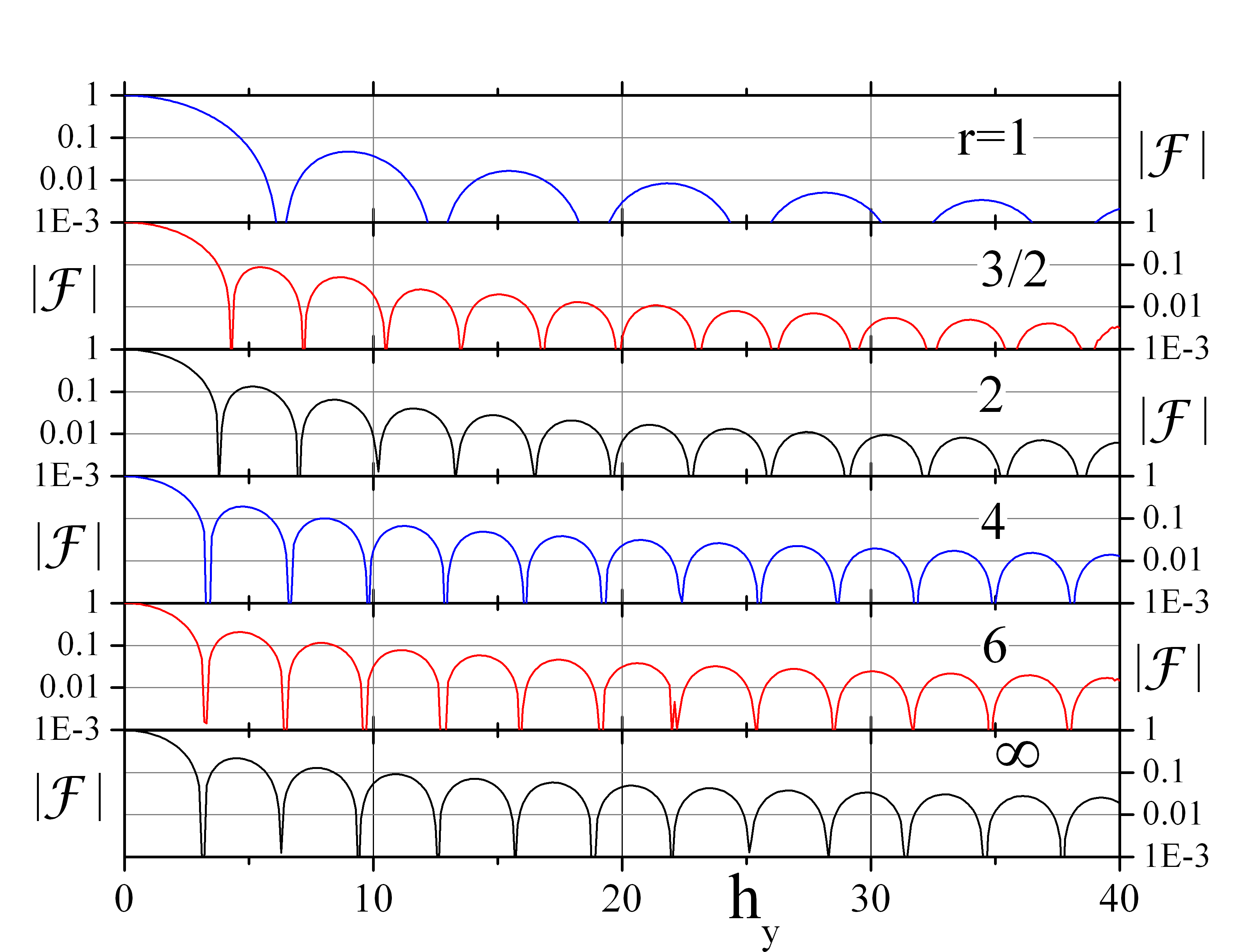}}
\caption{Characteristic area-independent functions, $\mathcal{F}_r$, of superelliptic JTJs for $r_x=r_y=r$ as a function of the normalized magnetic field, $h_y\equiv \kappa H a$, applied along the $Y$-axis: (a) for $r=1/4$, $1/3$, $1/2$, $2/3$, $3/4$ and $1$; (b) absolute value of the characteristic functions for $r=1$, $3/2$, $2$, $4$, $6$ and $\infty$.}
\label{patterns}
\end{figure}

\subsubsection*{$r=1/2$}
\vskip -8pt
$$\mathcal{F}_{1/2}(h_y)=\frac{12}{h_y^2}\left[ \sqrt{\frac{\pi h_y}{2}}\, \mathcal{S}\left(\sqrt{\frac{2h_y}{\pi}}\right) - \sin^2 \frac{h_y}{2} \right].$$


\subsubsection*{$r=2/3$: the astroid}
\vskip -8pt
$$\mathcal{F}_{2/3}(h_y)=\, _2F_3\left(\frac{5}{6},\frac{7}{6};\frac{4}{3},\frac{5}{3},2;-\frac{h_y^2}{4}\right).$$

\subsubsection*{$r=1$: the rhombus}
\vskip -8pt
\begin{equation}
\mathcal{F}_{1}(h_y)=\,_1F_2\left(1;\frac{3}{2},2;-\frac{h_y^2}{4}\right)=\frac{4 \sin^2\frac{h_y}{2}}{h_y^2}.
\label{rhombus}
\end{equation}
\vskip -4pt

\noindent More generally, for a small diamond-like JTJ of diagonals $2a$ and $2b$ in presence of a spatially homogeneous in-plane magnetic field of arbitrary orientations the characteristic function is \cite{nappi96}:

\vskip -8pt
\begin{equation}
\label{diamond}
\mathcal{F}^D(h_x,h_y)=2 \left(\frac{\cos h_x - \cos h_y} {h_y^2 - h_x^2} \right).
\end{equation}

\noindent Eq.(\ref{diamond}) reduces to Eq.(\ref{rhombus}) in the limit $h_x \to 0$.

\subsubsection*{$r=2$: the ellipse}
\vskip -8pt
\begin{equation}
\mathcal{F}_{2}(h_y)=\,_0F_1\left(\!;2;-\frac{h_y^2}{4} \right)=
2\frac{J_1(|h_y|)}{|h_y|}.
\label{ellipse}
\end{equation}
\vskip -4pt

\noindent In presence of a spatially homogeneous in-plane magnetic field $H$ of arbitrary orientations $\theta$ relative to the $Y$-axis, the characteristic function of a JTJ delimited by an ellipse of principal semi-axes $a$ and $b$ is \cite{ekin}:

\vskip -8pt
\begin{equation}
\label{IcEllipse}
\mathcal{F}^E[h(\theta)]= 2\frac{J_1(|h(\theta)|)}{|h(\theta)|},
\end{equation}


\noindent where $h(\theta)=\kappa H L(\theta)$ and $2L(\theta) \equiv 2 \sqrt{b^2 \cos^2 \theta + a^2 \sin^2 \theta}$ is the length of the projection of the junction in the direction normal to the externally applied magnetic field. Eq.(\ref{IcEllipse}) generalizes the so called \textit{Airy pattern} of a circular junction \cite{barone}; it was first reported by Peterson \textit{et al.} \cite{ekin} in 1990 and has been recently recalculated using a different approach \cite{JLTP16b}.

\subsubsection*{$r=3$}
\vskip -8pt
$$\mathcal{F}_{3}(h_y)=\!\,_0F_5\left(;\frac{1}{3},\frac{1}{2},\frac{5}{6},\frac{5}{6},\frac{4}{3};-\frac{h_y^6}{6^6}\right)+$$
\vskip -8pt
$$+\frac{ \Gamma\!\left(-\frac{1}{6}\right)h_y^2}{24 \sqrt[3]{2}\sqrt{\pi}\,\Gamma\!\left(\frac{4}{3}\right)}\,_1F_6\left(1;\frac{2}{3},\frac{5}{6}, \frac{7}{6},\frac{7}{6},\frac{4}{3},\frac{5}{3};-\frac{h_y^6}{6^6}\right)-\frac{\sqrt{\frac{\pi }{3}}\,\Gamma\!\left(-\frac{1}{6}\right)h_y^4}{972 \sqrt[3]{2}\, \Gamma\!\left(\frac{4}{3}\right)}\,_0F_5\left(;\frac{7}{6},\frac{3}{2}, \frac{3}{2}, \frac{5}{3},2;-\frac{h_y^6}{6^6}\right).$$

\subsubsection*{$r=4$: the rectellipse}
\vskip -8pt
$$\mathcal{F}_{4}(h_y)=\frac{4}{h_y^2}[\text{J}_2(h_y)+ \text{I}_2(h_y)]-\frac{\sqrt{\pi}\, \Gamma\!\left(\frac{3}{4}\right)h_y^2}{4 \Gamma\!\left(\frac{1}{4}\right)} \,_0F_3 \left(;\frac{5}{4},\frac{3}{2},2;\frac{h_y^4}{4^4}\right),$$

\noindent where $J_2$ and $I_2$ are, respectively, the second order regular and modified Bessel functions of the first kind.

\subsubsection*{$r=6$}
\vskip -8pt
$$\mathcal{F}_{6}(h_y)=\!\,_0F_5\left(\!;\frac{1}{3},\frac{1}{2},\frac{2}{3},\frac{5}{6},\frac{4}{3};-\frac{h_y^6}{6^6}\right)+\frac{\sqrt{\pi} h_y^4 \Gamma\!\left(\frac{5}{3}\right) \,_0F_5\left(\!;\frac{7}{6},\frac{4}{3},\frac{3}{2},\frac{5}{3},2; -\frac{h_y^6}{6^6}\right)} {144\,\times 2^{2/3} \Gamma\!\left(\frac{7}{6}\right)}-\frac{h_y^2 \,_0F_5\left(\!;\frac{2}{3}, \frac{5}{6},\frac{7}{6},\frac{4}{3},\frac{5}{3}; -\frac{h_y^6}{6^6}\right)}{4\,\times 2^{2/3}}.$$

\subsubsection*{$r \to \infty$: the rectangle}

$$\mathcal{F}_{\infty}(h_y) =\!\,_0F_1\left(\!;\frac{3}{2};-\frac{h_y^2}{4}\right)=\sqrt{ \frac{\pi}{2h_y}} J_{1/2}(h_y)=\Sinc(h_y)\equiv \frac{\sin\,h_y}{h_y}.$$

\noindent As expected, in the limit $r \to \infty$, the characteristic function tends to the \textit{sine cardinal} or \textit{Sinc} function, which yields the well-known Fraunhofer diffraction pattern of rectangular JTJs.
\vskip 8pt
\noindent Figs~\ref{patterns} (a) and (b) show how the critical-current pattern of a superelliptic junction gradually evolves for increasing values of the exponent $r$. For $r\leq 1$ the characteristic functions are drawn in Fig~\ref{patterns} (a) (we do not take the absolute value since they are positive functions). We observe that for $r<1/2$ the critical current decays monotonously; for larger values of the exponent some interference oscillations appear; however, since the characteristic functions do not have zeroes, we cannot strictly speak of sidelobes as far as $r<1$. The analysis of the closed form expressions for large field reveals that the MDPs decay as $1/h_y^{1+r}$ for $r\leq 1$. Fig~\ref{patterns} (b), for $r\geq1$, shows that, as $r$ increases, the pattern periodicity shifts for $2\pi$ to $\pi$ and the amplitudes of the close-in sidelobes steadily grow. At the same time the envelope asymptotic behavior for very large field is $\propto 1/h_y^{1+1/r}$ with the prefactors gradually decreasing from $4$ to $1$. Therefore, in the limit of $h_y \to \infty$, the power dependences of the modulation envelopes for $r$ and $1/r$ have the same scaling exponent, although different prefactors.

\section{The MDP of hybrid super\elli \Jos tunnel junctions} 

The characteristic functions of hybrid ($r_x=1/r_y$) superelliptic JTJs with the magnetic field applied along the $Y$-axis are reported below for some values of $r_x$ which permit analytical integration. For $r_x=1$ we recover the result of Eq.(\ref{rhombus}) for a rhomboidal JTJ, namely, $\mathcal{F}_{1,1}(h_y)=4 \sin^2(h_y/2)/h_y^2$.

\subsubsection*{$r_x=1/r_y=2$}
\vskip -8pt
$$\mathcal{F}_{2,1/2}(h_y)=15\left(\frac{3 \sin h_y}{h_y^5}-\frac{3 \cos h_y}{h_y^4}-\frac{ \sin h_y}{h_y^3}\right).$$
\vskip -4pt

\subsubsection*{$r_x=1/r_y=3$}
\vskip -8pt
\vskip -4pt
$$\mathcal{F}_{3,1/3}(\!h_y)\!=\!280\!\left[\!\frac{2240(\!1\!-\!\cos\!h_y)} {h_y^{10}}\!-\!\frac{2240\!\sin\!h_y}{h_y^9}\!+\!\frac{1120\!\cos\!h_y}{h_y^8}\!+\!\frac{360\! \sin\! h_y}{h_y^7}\!-\!\frac{80\!\cos\!h_y}{h_y^6}\!-\!\frac{12\!\sin\!h_y}{h_y^5}\! +\!\frac{\cos\! h_y\!-\!1/9}{h_y^4}\right].$$
\vskip -4pt

\subsubsection*{$r_x=1/r_y=4$}
\vskip -8pt
$$\mathcal{F}_{4,1/4}(h_y)=149175\left(-\frac{227026800}{h_y^{16}}+\frac{37837800}{h_y^{14}}-\frac{1871100}{h_y^{12}}+\frac{41580}{h_y^{10}}-\frac{455}{h_y^8}+\frac{2}{h_y^6}\right)\cos h_y +$$
\vskip -4pt
$$+9945\left(\frac{3405402000}{h_y^{17}}-\frac{1702701000}{h_y^{15}}+\frac{141579900}{h_y^{13}}-\frac{4573800}{h_y^{11}}+\frac{6825}{h_y^9}-\frac{525}{h_y^7}+\frac{1}{h_y^5}\right)\sin h_y.$$
\vskip -4pt
\begin{figure}[tb]
\centering
\subfigure[ ]{\includegraphics[width=7cm]{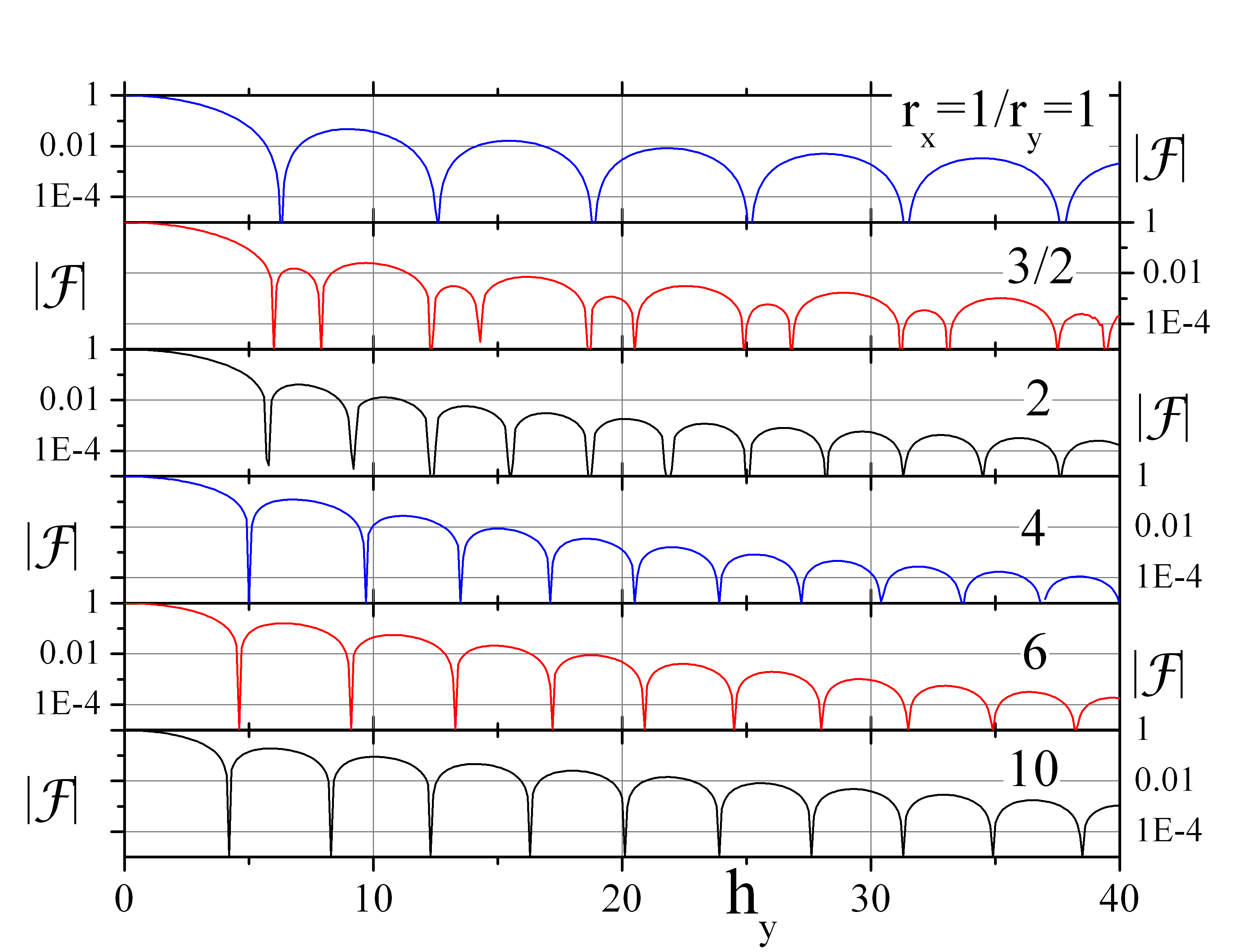}}
\subfigure[ ]{\includegraphics[width=7cm]{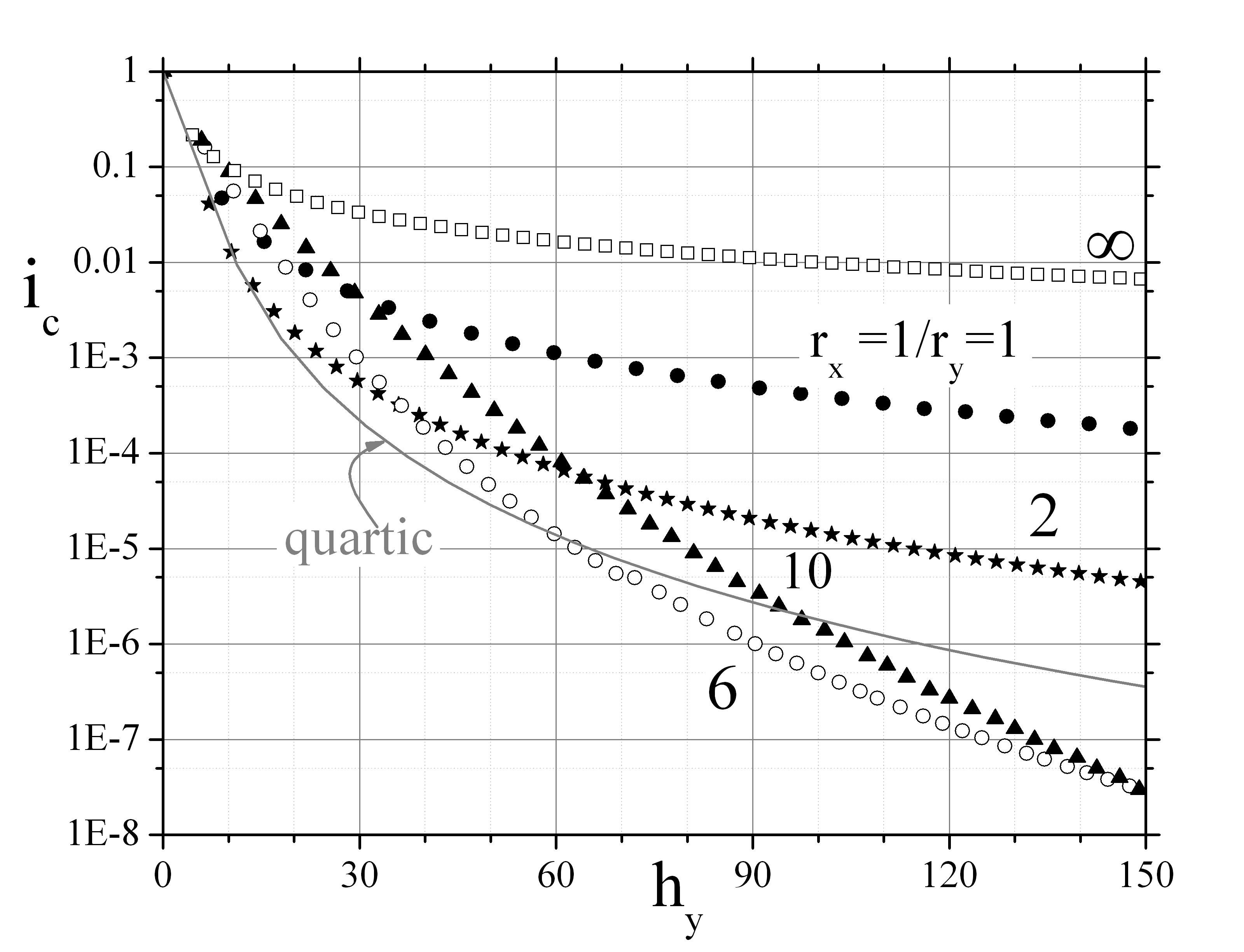}}
\caption{a) The absolute value of the characteristic functions, $\mathcal{F}_{r_x,r_y}$, of hybrid superelliptic JTJs for $r_x=1/r_y=1$, $3/2$, $2$, $4$, $6$ and $10$ as a function of the normalized magnetic field, $h_y\equiv \kappa H a$, applied along the $Y$-axis; b) the maxima of the first sidelobes relative to the central lobe for several values of the exponent $r_x$; the solid gray line refers to the envelope of the quartic characteristic function in Eq.(\ref{quartus}).}
\label{hybrpatterns}
\end{figure}

\noindent For larger (integer) exponents the characteristic functions are still calculable, but their formula become more and more complicated. Fig~\ref{patterns} (a) compares the characteristic functions for increasing values of exponent $r_x$ (from top to bottom). At variance with Fig~\ref{hybrpatterns} (b), now the vertical scale spans over five orders of magnitude, rather than three. It is important to stress that, by construction, all the junctions have the same transverse dimension $2a$ (no matter what it is, as far it is smaller than the Josephson penetration depth). At infinite the MDP envelopes asymptotically vanish as $1/h_y^{1+r_x}$, i.e., the roll-off is faster for larger exponents. However, the power law prefactors, $A_{r_x}$, drastically grow with the exponent; here few of them are reported: $A_1=4$, $A_2=15$, $A_3=280(1+1/9)\simeq 311$, $A_4=9945$, $A_5=576576(1+1/625)=577499$, $A_6=49579075$ and $A_{10}=48279601331512551$. Being the characteristic functions made up by the sum of many power law terms, the regime of fastest decay is achieved only for extremely large fields. Indeed, what really matter in applications is the required degree of critical-current suppression and the largest applicable magnetic field. In order to make unambiguous comparisons, it is therefore more appropriate to plot the amplitudes of the pattern oscillations versus the strength of the external field. In Fig~\ref{hybrpatterns} (b) the maxima of the first sidelobes relative to the central lobe are reported for some selected value of $r_x=1/r_y$. The solid gray line refers to the magnetic dependence of the envelope for the quartic junction \cite{peterson91} given in Eq.(\ref{quartic}), namely,
\vskip -8pt
\begin{equation}
\mathcal{F}_{quartic}(h_y)=60 \frac{2 h_y+ h_y \cos h_y - 3 \sin h_y}{h_y^5};
\label{quartus}
\end{equation}

\noindent it is seen that this expression asymptotically decays as $180/h_y^4$. We remark that a faster sidelobe suppression can be achieved with hybrid superellipses as long as $r_x=1/r_y>3$. 
Fig~\ref{hybrpatterns} (b) allows for a univocal comparison of the efficiency of a given magnetic field to suppress the junction critical current, since the same field normalization, i.e., the same transverse dimension, is considered. The diagram shows that the quartic shape is more efficient than any superelliptic shapes in most experimental situations where a current suppression of a factor $10^4$ is more than acceptable. In principle, larger sidelobe reductions can be more efficiently achieved by superelliptic junctions; however, theoretical arguments supported by experimental reports  \cite{houwman91} demonstrated that, due to lithographic limitations in  reproducing the very pointed extremity of lens-shaped junctions, a residual modulation persist in the magnetic pattern at extremely large field which decays as $1/h_y$.   

\section{Summary}

The physics of Josephson tunnel junctions drastically depends on their geometrical configurations. In this paper we have investigated small planar JTJs whose tunneling area are delimited by a superellipse or Lam\`e curve, a point symmetric plane figure that, in the most general case, is characterized by a pair of positive exponents. We have derived an integral expression (see Eq.(\ref{Frnot0Gen})) for the in-plane field dependence of their critical-currents for an arbitrary field direction. The characteristic function formalism has been used to deal with the pattern profiles which only depend on the junction shape and not on its dimensions. The characteristic functions have been expressed in term of generalized hypergeometric functions when the field is parallel to a symmetry axis. We have seen that a large variety of pattern profiles can be obtained as the junction shape is changed through the exponents. This collection of example shapes shows that for large fields the rate of sidelobe suppression follows a power law whose exponent results to be very large for the hybrid superellipses which are characterized by reciprocal exponents. 
%
%

%
%
%
%
%


\newpage

\begin{thebibliography}{99}

\bibitem{barone}  A. Barone and G. Patern\`o, {\em Physics and Applications of the Josephson Effect }(Wiley, New York, 1982).

\bibitem{ekin} R.L. Peterson and J.W. Ekin, {\it Physica C} {\bf 157}, 325 (1989); R.L. Peterson and J.W. Ekin, {\it Phys. Rev. B} {\bf 42}, 8014 (1990).

\bibitem{peterson91} R.L. Peterson, {\it Cryogenics} {\bf 31}, 132 (1991).

\bibitem{Dew-Hughes} D. Dew-Hughes, {\it Low Temp. Phys.} {\bf 27}, 713 (2001).

\bibitem{gijsbertsen95} J. G. Gijsbertsen, E. P. Houwman, B. B. G. Klopman, J. Flokstra, H. Rogalla, D. Quenter and S. Lemke,  {\it Physica C} {\bf 249}, 12 (1995).

\bibitem{kikuchi00} K. Kikuchi, H. Myoren, T. Iizuka, and S. Takada, {\it Appl. Phys. Letts.} {\bf 77}, 3660 (2000).

\bibitem{nappi96} C. Nappi, R. Cristiano, L. Frunzio, S. Pagano, and M. P. Lisitskii, {\it J. Appl. Phys.} {\bf 80}, 3401 (1996).

\bibitem{brian}  B. D. Josephson, {\it Phys. Lett.} {\bf 1}, 251 (1962).
%
\bibitem{wei} M. Weihnacht, {\it Phys. Stat Sol.} {\bf 32}, K169 (1969).

\bibitem{SUST13a} R. Monaco, V.P. Koshelets, A. Mukhortova, J. Mygind, {\it Supercond. Sci. Technol.} {\bf 26}, 055021 (2013).
%
\bibitem{owen} C.S. Owen and D.J. Scalapino, {\it Phys. Rev.} {\bf 164}, 538 (1967).

\bibitem{PRB12} R. Monaco, J. Mygind, and V.P. Koshelets, {\it Phys. Rev. B} {\bf 85}, 094514 (2012).

\bibitem{PhysC16b} R. Monaco, {\it Physica C} {\bf 525-526}, 52 (2016).

\bibitem{wolfram} \href{url}{http://mathworld.wolfram.com/Superellipse.html}

\bibitem{JLTP16b} R. Monaco, {\it J. Low Temp. Phys.} {\bf 184}, 979 (2016).





\bibitem{houwman91} E. P. Houwman, J. G. Gijsbertsen, J. Flokstra and H. Rogalla,  {\it Physica C} {\bf 183}, 339 (1991).



%
%
%
%
%
%
%
%
%














%
%
%
%
%
%
%
%
%
%
%



















 





















%






























\end{thebibliography}
\end{document}